\documentclass[journal,twoside,web]{ieeecolor}
\usepackage{generic}
\usepackage{cite}
\usepackage{amsmath,amssymb,amsfonts}
\usepackage{algorithmic}
\usepackage{graphicx}
\usepackage{textcomp}
\def\BibTeX{{\rm B\kern-.05em{\sc i\kern-.025em b}\kern-.08em
		T\kern-.1667em\lower.7ex\hbox{E}\kern-.125emX}}
\begin{document}
\title{Task-Space Control of Robot Manipulators based on Visual SLAM}
	
\author{Seyed Hamed Hashemi, Jouni Mattila
\thanks{© 2022 IEEE. Personal use of this material is permitted. Permission from IEEE must be obtained for all other uses, including reprinting/republishing this material for advertising or promotional purposes, collecting new collected works for resale or redistribution to servers or lists, or reuse of any copyrighted component of this work in other works. This work has been submitted to the IEEE for possible publication. Copyright may be transferred without notice, after which this version may no longer be accessible.}
\thanks{This work was supported by the Academy of Finland as part of the “High-precision autonomous mobile manipulators for future digitalized construction sites” project [Grant No. 335569].}
\thanks{The authors are with the Unit of Automation Technology and Mechanical Engineering, Faculty of Engineering and Natural Sciences, Tampere University, Tampere, Finland (e-mail: hamed.hashemi@tuni.fi, jouni.mattila@tuni.fi).}} 
	
\maketitle
	
\begin{abstract}
This paper aims to address the open problem of designing a globally stable vision-based controller for robot manipulators. Accordingly, based on a hybrid mechanism, this paper proposes a novel task-space control law attained by taking the gradient of a potential function in $SE(3)$. The key idea is to employ the Visual Simultaneous Localization and Mapping (VSLAM) algorithm to estimate a robot pose. The estimated robot pose is then used in the proposed hybrid controller as feedback information. Invoking Barbalat’s lemma and Lyapunov's stability theorem, it is guaranteed that the resulting closed-loop system is globally asymptotically stable, which is the main accomplishment of the proposed structure. Simulation studies are conducted on a six degrees of freedom (6-DOF) robot manipulator to demonstrate the effectiveness and validate the performance of the proposed VSLAM-based control scheme.
\end{abstract}
	
\begin{IEEEkeywords}
Hybrid systems, robot manipulator, task-space control, visual SLAM
\end{IEEEkeywords}
	
\section{Introduction}
\IEEEPARstart{T}{he} control problem of robot manipulators has been widely investigated in significant research from recent decades \cite{lynch2017modern}. This is because robotic systems have a wide range of applications, from agriculture to surgery, industry, and the military \cite{wang2019dynamic}. Existing control strategies for robot manipulators can be categorized into two types: 1) task-space control methods and 2) joint-space control methods. The former are designed by defining trajectory-tracking errors inside task-space, and afterwards, transforming the control law into torque space. In the latter by contrast, the end-effector trajectory is mapped into joint space through inverse kinematics to determine the path-tracking errors \cite{chen2020model}.
	
Since most robot tasks in industry are defined in end-effector space, much research has focused on designing controllers directly in task-space \cite{hu2022task}. The first task-space controller was introduced in \cite{10.1115/1.3139651}, where it was argued that the presented approach is applicable even in the case of singularity or redundancy. Slotine and Li in \cite{slotine1987adaptive} introduced a feedback law which takes advantage of both a Proportional–Derivative-type (PD-type) controller and a full dynamics feed-forward compensator. This controller eliminates the need for feedback information on joint accelerations and inverse of the estimated inertia matrix. The work in \cite{slotine1987adaptive} was revisited by \cite{loria2005uniform}, where it was proven that a system that tracks desired paths while keeping the regressor matrix exciting over time satisfies the necessary and sufficient conditions of uniform global asymptotic stability. Since the demand for robot manipulators has grown remarkably in different industrial applications, numerous recent studies have been devoted to designing various controllers in task-space, such as quaternion-based \cite{xian2004task}, dual quaternion-based \cite{huang2022ude}, and inverse dynamics controllers \cite{obregon2021predefined}. 

The vision-based controller is another commonly used method for solving the robot control problem in Cartesian space. It utilizes camera measurements as feedback information. This is because robots endowed with cameras can provide more environmental information and are more flexible in dealing with unstructured environments \cite{jin2021policy}. Based on the class of feedback information provided by vision sensors, vision-based controllers can be categorized into three main types \cite{bechlioulis2019robust}: 1) image-based visual servoing (IBVS), 2) position-based visual servoing (PBVS), and 3) hybrid visual servoing (HVS). For its advantages, such as ease of implementation, robustness against image noise, and camera calibration, 2D IBVS has received considerable attention. Nonetheless, IBVS still suffers from some challenges: 1) it can only ensure stability in a region close enough to the desired set, 2) it requires an exact interaction matrix, 3) it can get stuck in a local minima, and 4) it requires all features to be tracked in two consecutive frames \cite{keshmiri2016image}. Accordingly, PBVS has been widely researched since it provides 3D pose estimation, a broad field of view, and higher accuracy for motion control. Nevertheless, this method has its own disadvantages, such as sensitivity to camera calibration and a need for accurate robot and camera models \cite{chen2021auto}. Moreover, the main shortcoming of existing vision-based control techniques is that they still do not consider the effects of robot dynamics in stability proofs.
	
In light of the aforementioned discussion, this paper introduces a new control scheme that exploits every advantage of the vision-based control strategy while trying to overcome its problems. To the best of the authors' knowledge, the stability of state-of-the-art vision-based control methods is only valid locally in confined task-space. Furthermore, the stability of these techniques depends on the number of visible features and camera calibration. Consequently, for the first time, this paper suggests utilizing a VSLAM observer to estimate robot pose because its stability is independent of camera and feature information. As a result, this paper proposes a new VSLAM-based control structure that, to improve upon  existing task-space control approaches, makes the following contributions:
	
$\bullet$ A new hybrid feedback law is designed directly in task-space to control end-effector pose. The proposed control law is a gradient-based controller obtained by taking the gradient of a positive-valued continuously differentiable function in $SE(3)$.
	
$\bullet$ A geometric VSLAM algorithm is derived directly from the Lie group of $SLAM_n(3)$. The introduced VSLAM observer estimates the robot pose, which is then used as task-space feedback information for the proposed controller. 
	
$\bullet$ The main contribution of this paper is to show how a VSLAM-based control structure gains global asymptotic stability by incorporating robot dynamics. 
	
This paper consists of six sections, including the introduction. Preliminary mathematical definitions, a generic background on hybrid systems, SLAM kinematics, and the kinematic and dynamic equations of the robot manipulator are provided in section 2. A visual SLAM algorithm is described in section 3. Section 4 presents a design and stability analysis of the proposed hybrid feedback law. Section 5 demonstrates the performance of the proposed VSLAM-based control structure based on simulation results. By providing some concluding remarks, section 6 summarizes the paper.
	
\section{Preliminaries }
\subsection{Notation}
By $\mathbb{R}$, this paper denotes the set of real numbers, and by $\mathbb{N}$, it denotes the set of natural numbers. $\mathbb{S}^n:=\{y\in \mathbb{R}^{n+1}: \|y\|=1 \} $ and $\mathbb{B} :=\{y\in \mathbb{R}^n: \|y\| \le1 \}$, respectively, represent a unit $n$-dimensional sphere and a closed unit $n$-dimensional ball. ${{\mathbb{R}}^{n}}$ denotes $n$-dimensional Euclidean space, where $\{e_i\}_{1\le i \le n} \subset {{\mathbb{R}}^{n}}$ is the canonical basis of ${{\mathbb{R}}^{n}}$. The Euclidean norm of a vector $x \in {{\mathbb{R}}^{n}}$ is given by $\left\| x \right\|=\sqrt{\left\langle x,x \right\rangle }$, where $\left\langle x,y \right\rangle :={{x}^{T}}y$ are the inner products and ${{\left\| x \right\|}_{\mathcal{A}}}:={{\min }_{y\in \mathcal{A}}}\left\| x-y \right\|$. Given a matrix $A \in \mathbb{R}^{n \times n}$, its trace, determinant, transpose, and skew-symmetric parts and Frobenius norm are defined as $\text{tr}(A)$, $\text{det}(A)$, $A^T$, $\text{skew}(A)=(A-A^{T})/2$, and  $\left\| A \right\|_F=\sqrt{\left\langle A,A \right\rangle}=\sqrt{\text{tr}(A^{T}A)}$, respectively. In this paper, matrix $R$ belongs to the special orthogonal group of order three $SO(3):=\{R \in \mathbb{R}^{3\times3}:R^TR=RR^T=I, \text{det}(R)=1 \}$, which denotes the attitude of a rigid body. The Lie algebra of $SO(3)$ is denoted by $\mathfrak{so}(3)=\{A \in \mathbb{R}^{3 \times 3}: A^T=-A \}$. The matrix Lie group $SLAM_n(3):=\{ \mathcal{X}= \Psi(R,p,\eta):R \in SO(3), p \in \mathbb{R}^3, \eta \in \mathbb{R}^{3\times n}\}$ is the $SLAM$ group and  $ {\mathcal{X}}^{-1} = \Psi (R^T,-R^Tp,-R^T \eta)$ is its inverse. The Lie algebra of $SLAM_n(3)$ is given by
	
\begin{equation*}\label{eq1}
\begin{split}
&\mathfrak{slam}_n(3):=\{\mathcal{V}(\omega,v,\xi)= \left[\begin{array}{c c|c} 
\Gamma(\omega) & v & \xi \\
\hline 
0_{n+1\times3} & 0_{n+1\times1} & 0_{n+1 \times n} 
\end{array}\right]\\
& :\omega, v \in \mathbb{R}^3, \xi \in \mathbb{R}^{3\times n} \}.\\
\end{split}
\end{equation*}
	
Let $\mathcal{M}$ be a smooth manifold and $T_{\mathcal{X}} \mathcal{M}$ denote its tangent space, as in $T_{\mathcal{X}}SLAM_n(3):= \{\mathcal{X} \mathcal{V}: \mathcal{X} \in SLAM_n(3) \ \text{and} \ \mathcal{V} \in \mathfrak{slam}_n(3)\}$. The map $Ad_{\mathcal{X}}:SLAM_n(3) \times \mathfrak{slam}_n(3) \rightarrow \mathfrak{slam}_n(3)$ is called the adjoint map for the $SLAM$ group and its associated Lie algebra $\mathfrak{slam}_n(3)$, which transforms a tangent vector of one element into a tangent vector of another. The following maps are frequently utilized throughout the paper.
	
\begin{equation}\label{eq2}
	\begin{split}
		&\Gamma (y)=\left[ \begin{matrix}
			0 & -{{y}_{3}} & {{y}_{2}}  \\
			{{y}_{3}} & 0 & -{{y}_{1}}  \\
			-{{y}_{2}} & {{y}_{1}} & 0  \\
		\end{matrix} \right], \\
		& \varphi(A)=\frac{1}{2} \left[ \begin{matrix}
			A_{(3,2)}-A_{(2,3)} \\
			A_{(1,3)}-A_{(3,1)} \\
			A_{(2,1)}-A_{(1,2)} \\
		\end{matrix} \right], \ \bar{\varphi}(\mathbb{A})= \left[ \begin{matrix}
			\varphi(A) \\
			\frac{1}{2} y \\
		\end{matrix} \right], \\
		&\Psi(R,p,\eta)= \left[\begin{array}{c c|c} 
			R & p & \eta \\
			0_{1\times3} & 1 & 0_{1\times n} \\
			\hline 
			0_{n\times3} & 0_{n\times1} & I_{n \times n} 
		\end{array}\right], \\
		&\bar{\Psi}(R,p)= \left[\begin{array}{c c} 
			R & p \\
			0_{1\times3} & 1\\
		\end{array}\right], \ \mathbb{A}= \left[\begin{array}{c c} 
			A & y \\
			c & d\\ 
		\end{array}\right], \\
		&\Upsilon(B)=\Upsilon(\left[ \begin{matrix}
			A & B_2 \\
			B_3^T & B_4 \\
		\end{matrix} \right])= \left[ \begin{matrix}
			\text{skew}(A) & B_2 \\
			0_{n+1 \times 3} & 0_{n+1 \times n+1} \\
		\end{matrix} \right],\\
		& \bar{\Upsilon}(\mathbb{A})=\Upsilon(\left[ \begin{matrix}
			\text{skew}(A) & y \\
			0_{1\times3} & 0\\
		\end{matrix} \right]), \\
		& AD_{\mathbb{X}} =\left[ \begin{matrix}
			R & 0_{3 \times 3} \\
			\Gamma(p)R & R\\
		\end{matrix} \right], \ \forall \mathbb{X} = \bar{\Psi}(R,p)\\
		& y \in \mathbb{R}^{3 \times 1}, A \in \mathbb{R}^{3 \times3} , c \in \mathbb{R}^{1 \times 3}, \ d \in \mathbb{R}\\
		& (B_2,B_3) \in \mathbb{R}^{3 \times n+1}, B_4 \in \mathbb{R}^{n+1 \times n+1} \\
	\end{split}
\end{equation}
	
Moreover, $\nabla_{\mathcal{X}}m \in T_{\mathcal{X}} \mathcal{M}$ represents the gradient of a differentiable smooth function $m:\mathcal{M} \to \mathbb{R}$, which is determined throuh the following equation:
	
\begin{equation}\label{eq3}
	dm.\mathcal{X}\mathcal{V}= \left\langle \nabla_{\mathcal{X}}m,\mathcal{X}\mathcal{V} \right\rangle_\mathcal{X}=\left\langle \mathcal{X}^{-1} \nabla_{\mathcal{X}}m,\mathcal{V} \right\rangle
\end{equation}
	
In Equation (\ref{eq3}), $dm$ and $\left\langle.,.\right\rangle_\mathcal{X}$ stand for the differential of $m$ and a Riemannian metric on the matrix Lie group, respectively, such that
	
\begin{equation*}\label{eq4}
	\left\langle \mathcal{X}\mathcal{V}_1,\mathcal{X}\mathcal{V}_2 \right\rangle_\mathcal{X} = \left\langle \mathcal{V}_1,\mathcal{V}_2 \right\rangle.
\end{equation*}
	
A Rodrigues formula $\Re :\mathbb{R}\times {\mathbb{S}^{2}}\to SO(3)$ defined by
	
\begin{equation}\label{eq5}
	\begin{split}
		&\Re (\theta ,y)=I+\sin (\theta )\Gamma (y)+(1-\cos (\theta )){{\Gamma }^{2}}(y), \quad \text{or} \\
		&\Re (\theta ,y)=\exp(\theta \Gamma(y)),\\
	\end{split}
\end{equation}
is utilized to describe a rotation matrix $R \in SO(3)$ in terms of its axis $y\in {\mathbb{S}^{2}}$ and angle $\theta \in \mathbb{R}$ of rotation.
	
\subsection{Hybrid System Framework}
The following equation describes the framework for hybrid dynamical systems $\mathcal{H}$ used throughout this paper \cite{hashemi2021quaternion}:
	
\begin{equation}\label{eq6}
	\mathcal{H}:\left\{ \begin{matrix}
		\dot{x}=f(x,u), & (x,u)\in C \\
		{{x}^{+}}=g(x,u), & (x,u)\in D \\
	\end{matrix} \right.
\end{equation} 
	
Here, $x \in \mathbb{R}^n$ and $u \in \mathbb{R}^m$ denote the state vector and the input of $\mathcal{H}$, respectively. The flow map $f:{{\mathbb{R}}^{n}}\times {{\mathbb{R}}^{m}}\to {{\mathbb{R}}^{n}}$ defines the continuous evolution of $x$ when $(x,u)$ belongs to the flow set $C\subset {{\mathbb{R}}^{n}}\times {{\mathbb{R}}^{m}}$. The jump map $g:{{\mathbb{R}}^{n}}\times {{\mathbb{R}}^{m}}\to {{\mathbb{R}}^{n}}$ describes the behavior of the systems during jumps when $(x,u)$ belongs to the jump set $D\subset {{\mathbb{R}}^{n}}\times {{\mathbb{R}}^{m}}$. $C$ and $D$ illustrate where continuous evolution and jumps are permitted, respectively. A solution to $\mathcal{H}$ is defined in a hybrid time domain $E \subset {{\mathbb{R}}_{\ge 0}}\times \mathbb{N}$, which is parameterized by the time variable $t \in \mathbb{R}_{\ge 0}$ and jump variable $j \in \mathbb{N}$. The subset $E$ is a hybrid time domain if it can be written as $E=\bigcup\limits_{i=1}^{I}{(\left[ {{t}_{i}},{{t}_{i+1}} \right],i)}$ for finite sequences of time $0={{t}_{0}}\le {{t}_{1}}\cdots \le {{t}_{I+1}}$.
	
\textbf{Lemma 1} \cite{teel2012lyapunov}: For the hybrid system $\mathcal{H}$, the closed set $\mathcal{A} \subset \mathbb{R}^n$ is judged locally exponentially stable if there exist $(\alpha_1 > \alpha_2, s_1, s_2, n)\in \mathbb{R}_{\ge0}$ and a continuously differentiable function $V:\mathbb{R}^n \to \mathbb{R}_{\ge0}$ such that the following inequalities hold:
	
\begin{equation}\label{eq7}
	\begin{split}
		& \alpha_2{\left\| x \right\|}_{\mathcal{A}}^n \le V(x) \le \alpha_1{\left\| x \right\|}_{\mathcal{A}}^n, \\
		& \forall x \in (C \cup D \cup g(D)) \cap (\mathcal{A}+s_1\mathbb{B})\\
		&\left\langle \nabla V(x),f \right\rangle \le -s_2V(x), \quad \forall x \in C \cap (\mathcal{A}+s_1\mathbb{B})\\
		& V(g) \le \exp(-s_2)V(x), \quad \forall x \in D \cap (\mathcal{A}+s_1\mathbb{B}).\\
	\end{split}
\end{equation}
	
Function $V$ is defined in an open set containing the closure of $C$. When $s_1 \to \infty$ and both $s_2 \to 0$ and $s_1 \to \infty$, set $\mathcal{A}$ is said to be globally exponentially stable and globally asymptotically stable, respectively.
	
\subsection{SLAM Kinematics}
The kinematic equations of motion of a rigid body and the $i^{th}$ landmark can be expressed as
	
\begin{equation}\label{eq8}
	\dot{R}=R\Gamma (\omega ),
\end{equation}
\begin{equation}\label{eq9}
	\dot{p}=Rv,
\end{equation}
\begin{equation}\label{eq10}
	\quad \quad	\quad \dot{\eta_i}=R\xi_i, \quad i=1,\dots,n
\end{equation}
where $\xi_i \in \mathbb{R}^3$, $\omega \in \mathbb{R}^3$, and $v \in \mathbb{R}^3$, respectively, denote the linear speed of the $i^{th}$ landmark, angular rate, and linear velocity of a rigid body with respect to the body-fixed frame $\mathcal{B}$. Furthermore, $p \in \mathbb{R}^3$ represents the position of a rigid body in the inertial frame $\mathcal{I}$ and $\eta_i \in \mathbb{R}^3$ represents the location of the $i^{th}$ landmark in $\mathcal{I}$. The motion kinematics in (\ref{eq6}-\ref{eq8}) can be rephrased more compactly as
	
\begin{equation}\label{eq11}
	\dot{\mathcal{X}}=\mathcal{X}\mathcal{V}.
\end{equation}
	
This paper focuses on stationary landmarks, which signify that $\xi_i=0$. The robot is equipped with sensors to measure its linear and angular velocity. The robot is also equipped with a camera that can measure ranges $\theta_b=\|\eta_i-p\|$ and bearings $\jmath=R^T(\eta_i-p)/\theta_b$ in relation to landmarks. Accordingly, $\beta_i$ denotes a camera measurement that contains both range and bearing measurements and is given by
	
\begin{equation}\label{eq12}
	\beta_i:=\mathcal{X}^{-1}r_i=\left[\begin{array}{c} 
		R^T(\eta_i-p) \\
		1 \\
		-e_i 
	\end{array}\right], \ r_i=\left[\begin{array}{c} 
		0_{3\times1} \\
		1 \\
		-e_i 
	\end{array}\right]. 
\end{equation}
	
\subsection{Robot Manipulator Dynamics} 
The dynamics of an $n$-link robot manipulator can be governed by the following so-called Euler–Lagrange equation \cite{9965565}:
	
\begin{equation}\label{eq13}
	M(q) \ddot{q}+C(q,\dot{q})\dot{q}+G(q)+F(\dot{q}) = \tau,
\end{equation}
where $q = [q_1,q_2, \cdots, q_n]^T \in \mathbb{R}^n$ is the joint position and $\dot{q} = [\dot{q_1},\dot{q_2}, \cdots, \dot{q_n}]^T \in \mathbb{R}^n$ is the joint velocity. Furthermore, $M(q) \in \mathbb{R}^{n \times n}$ is the nominal inertia matrix, $C(q,\dot{q}) \in \mathbb{R}^{n \times n}$ represents the nominal Coriolis-centrifugal matrix, $G(q) \in \mathbb{R}^n$ is the gravity vector, and $F(\dot{q}) \in \mathbb{R}^n$ contains the frictional force coefficients. Moreover, $\tau \in \mathbb{R}^n$ represents the applied control torque. The forward kinematics of a robotic manipulator provide a map between Cartesian space and joint space given by
	
\begin{equation}\label{eq14}
	x_{end}(t) = h(q(t)).
\end{equation}
	
Here, $x_{end}$ denotes the attitude and position of a manipulator end-effector in task-space. Taking the derivative with respect to time on both sides of (\ref{eq14}) yields the relation between the joint and Cartesian space velocities.
	
\begin{equation}\label{eq15}
	\dot{x}_{end}(t) = \frac{\partial h(q)}{\partial q}\dot{q}(t) = J\dot{q}(t), \ \text{or}, \ \mathcal{Z} = \left[ \begin{matrix}
		\omega \\
		v  \\
	\end{matrix} \right] = J\dot{q}(t) \\
\end{equation}
	
Here, $J$ is the Jacobian matrix of the forward kinematics, and $\omega,v$ represent the angular and linear velocity components of the end-effector velocity vector, respectively.
	
\section{Visual SLAM Algorithm}
This section presents the visual SLAM algorithm, which is used to estimate end-effector poses. There is an extensive literature full of various methods for solving the VSLAM problem. Consequently, existing VSLAM algorithms fall into three categories: optimization-based methods \cite{mur2015orb}, geometric-type techniques \cite{mahony2021homogeneous}, and Kalman-type algorithms \cite{zhou2015structslam}. All these techniques have their own advantages and disadvantages. For example, geometric-type algorithms can only guarantee almost global stability due to the existence of sets with Lebesgue measure zero in $SO(3)$. Likewise, Kalman-type methods and optimization-based strategies suffer from performance dependency on the initialization and also cannot ensure stability. Consequently, this paper makes use of the VSLAM algorithm introduced by the authors in \cite{hashemi2022global}, where it was proven that the provided VSLAM method can guarantee global asymptotic stability and overcome problems associated with existing VSLAM algorithms. 
	
The first step in designing the proposed VSLAM algorithm is defining the potential function, $\mathcal{U}: SLAM_n(3) \rightarrow \mathbb{R}$, which is given by
	
\begin{equation}\label{eq16}
	\begin{split}
		& \mathcal{U}(\mathcal{X})=\frac{1}{2}\text{tr}((I-\mathcal{X})A(I-\mathcal{X})^T), \\
		& A:=\sum_{i=1}^{n} k_ir_i{r_i}^T, \quad k_i \in \mathbb{R}_{\ge0}. \\
	\end{split}
\end{equation}
	
The following identity is useful in determining the gradient of potential function $\nabla_{\mathcal{X}} \mathcal{U}$, which is obtained with the aid of a Riemannian metric in $SLAM_n(3)$ and identities in the Appendix.
	
\begin{equation}\label{eq17}
	\begin{split}
		& d\mathcal{U}.\mathcal{X}\mathcal{V}=\left\langle \mathcal{X}^{-1} \nabla_{\mathcal{X}}\mathcal{U},\mathcal{V} \right\rangle, \quad \text{or} ,\\
		& d\mathcal{U}.\mathcal{X}\mathcal{V}=\text{tr}(-A(I-\mathcal{X})^T\mathcal{X}\mathcal{V}) \\
		& =\left\langle \Upsilon(\mathcal{X}^{-1}(\mathcal{X}-I)A),\mathcal{V} \right\rangle =\left\langle \Upsilon((I-\mathcal{X}^{-1})A),\mathcal{V} \right\rangle \\
	\end{split}
\end{equation}
	
As a result, the gradient of potential function $\mathcal{U}$ with respect to $\mathcal{X}$ is determined as follows:
	
\begin{equation}\label{eq18}
	\nabla_{\mathcal{X}}(\mathcal{U})=\mathcal{X}\Upsilon((I-\mathcal{X}^{-1})A)
\end{equation}
	
Estimation error is often defined as the difference between the true state value $\mathcal{X}$ and estimated state value $\hat{\mathcal{X}}$, i.e., $\tilde{\mathcal{X}}=\mathcal{X}\hat{\mathcal{X}}^{-1}$ with $\tilde{R}=R\hat{R}^T$, $\tilde{p}=p-\tilde{R}\hat{p}$, and $\tilde{\eta}=\eta-\tilde{R}\hat{\eta}$. Hence, the following identities represent the potential function (\ref{eq16}) and its gradient (\ref{eq18}) in terms of estimation error:
	
\begin{equation}\label{eq19}
	\begin{split}
		& \Upsilon(\sum_{i=1}^{n} k_i(r_i-\hat{\mathcal{X}}\beta_i)r_i^T)=\Upsilon((I-\tilde{\mathcal{X}}^{-1})A), \ (a) \\
		& \sum_{i=1}^{n} k_i\|r_i-\hat{\mathcal{X}}\beta_i\|^2=\text{tr}((I-\tilde{\mathcal{X}})A(I-\tilde{\mathcal{X}})^T), \ (b) \\
	\end{split}
\end{equation}
	
The following equation describes the dynamics of the VSLAM algorithm introduced by the authors in \cite{hashemi2022global}:
	
\begin{equation}\label{eq20}
	\begin{split}
		&\begin{cases}
			\dot{\hat{\mathcal{X}}}=\hat{\mathcal{X}}(\mathcal{V}-\Delta), & \hat{\mathcal{X}} \in C \\
			\dot{q}=0, \\
		\end{cases}\\ 
		&\begin{cases}
			\hat{\mathcal{X}}^+=\mathcal{X}_q, & \hat{\mathcal{X}} \in D \\
			q^+=\underset{q \in \mathcal{Q}}{\arg\min} \ \mathcal{U}(\tilde{\mathcal{X}_q}),
		\end{cases}\\
		& C:=\{(\mathcal{U}(\tilde{\mathcal{X}})-\min_{\tilde{\mathcal{X}_q}\in \mathcal{Q}}  \mathcal{U}(\tilde{\mathcal{X}_q})\le \delta), \\
		& D:=\{(\mathcal{U}(\tilde{\mathcal{X}})-\min_{\tilde{\mathcal{X}_q}\in \mathcal{Q}} \mathcal{U}(\tilde{\mathcal{X}_q})\ge \delta), \\
		&\mathcal{X}_q = \Psi(\Re(q\theta,\ell),0,0) \Psi(\hat{R},\hat{p},\hat{\eta}), \quad q \in \mathbb{N}\\
		&\Delta=-Ad_{\hat{\mathcal{X}}^{-1}}\Upsilon(\sum_{i=1}^{n} k_i(r_i-\hat{\mathcal{X}}\beta_i)r_i^T)K,\\
	\end{split}
\end{equation}
	
Here, $K:=k_o I_{n+4 \times n+4}$ with $k_o \in \mathbb{R}_{>0}$ is the observer gain, $q \in \mathbb{N}$ belongs to a compact set $\mathcal{Q}=\{\mathcal{X}_q \in SLAM_n(3): q \in \mathbb{N}, \ell \in \mathbb{S}^2, \theta \in \mathbb{R}_{>0} \}$, and $\tilde{\mathcal{X}_q}=\mathcal{X}\mathcal{X}_q^{-1}$. Moreover, $(\theta,\delta) \in \mathbb{R}_{>0}$, and $\ell \in \mathbb{S}^2$ are arbitrary constants and an arbitrary fixed vector, respectively.
	
\textbf{Theorem 1}: Consider the SLAM kinematics (\ref{eq11}) evolving on $SLAM_n(3)$ along with bounded measurements (\ref{eq12}). The VSLAM algorithm defined by (\ref{eq20}) is a global asymptotic convergent observer, i.e., state estimation error $\tilde{\mathcal{X}}$ globally asymptotically converges to $I_{n+4 \times n+4}$. 
	
\textbf{Proof}: The proof of Theorem 1 is omitted here; however, full details can be found in \cite{hashemi2022global}.
	
\section{Proposed Hybrid Feedback Law}
The proposed hybrid feedback law is developed in this section. In the past decade, hybrid controllers have frequently been used for stabilizing systems evolving in matrix Lie groups \cite{hashemi2022observer} since it was proven that continuous and discontinuous feedback laws cannot globally stabilize these systems in the desired set \cite{hashemi2021global}. This is due to the non-contractibility of the configuration space of the attitude and existence of sets that have Lebesgue measure zero \cite{hashemi2021observer}. Nonetheless, to the best of the authors' knowledge, hybrid controllers have not yet been applied to a robot manipulator. Consequently, for the first time, this paper designs a new hybrid feedback law to control robot manipulators in task-space. The proposed hybrid feedback law is directly designed in $SE(3)$ (end-effector configuration space) by constructing a potential function in $SE(3)$ and taking its gradient. Accordingly, potential function $\mathbb{U}: SE(3) \times \mathbb{R} \rightarrow \mathbb{R}$ is defined as follows:
	
\begin{equation}\label{eq21}
	\begin{split}
		&\mathbb{U}(\mathbb{X},h)=\frac{1}{2}\text{tr}((I-\mathbb{X}_h\mathbb{X})G(I-\mathbb{X}_h\mathbb{X})^T), \\
		& \mathbb{X}_h = \left[ \begin{matrix}
			\Re (\theta_h ,\jmath) & 0_{3 \times 1} \\
			0_{1\times3} & 1\\
		\end{matrix} \right] \\
	\end{split}
\end{equation}
	
Here, $G \in \mathbb{R}^{4 \times 4} $ is a symmetric positive definite matrix, $\mathbb{X} \in SE(3):=\{ \mathbb{X}= \bar{\Psi}(R,p):R \in SO(3), p \in \mathbb{R}^3\}$, $\jmath \in \mathbb{S}^2$ is an arbitrary constant vector, and $\theta_h \in \mathbb{R}$ belongs to a compact set $\Xi := \{\theta_h \in \mathbb{R}: |\theta_h| \le \pi/2 \}$. By following the same procedure of gradient calculation for the previous potential function, the gradient of $\mathbb{U}(\mathbb{X},h)$ is determined as follows (for details, see \cite{wang2021hybrid}):
	
\begin{equation}\label{eq22}
	\bar{\varphi}(\mathbb{X}^{-1}\nabla_{\mathbb{X}}\mathbb{U}(\mathbb{X},h))=AD_{\mathbb{X}_h}^{-T}\bar{\varphi}((I-(\mathbb{X}_h\mathbb{X})^{-1})G)
\end{equation}
	
Consequently, the proposed hybrid feedback control law is defined as
	
\begin{equation}\label{eq23}
	\begin{split}
		& \tau^* = N(q,\dot{q})-M(q)J^{-1}(\dot{J}\dot{q}+\bar{\varphi}(\mathbb{X}_e^{-1}\nabla_{\mathbb{X}_e}\mathbb{U}(\mathbb{X}_e,h)+G_d\mathbb{Y})), \\
		& N(q,\dot{q}) = C(q,\dot{q})\dot{q}+G(q)+F(\dot{q}). \\
	\end{split}
\end{equation}
	
In Equation (\ref{eq23}), $G_d=g_dI_{6\times6}$ with $g_d \in \mathbb{R}_{>0}$ is the controller gain, and definitions of $\mathbb{X}_e$ and $\mathbb{Y}$ are given in the proof of Theorem 2. The following hybrid mechanism calculates the switching variable $h$:
	
\begin{equation}\label{eq24}
	\begin{split}
		&\begin{cases}
			\dot{h}=0, & (\mathbb{X}_e,h) \in C' \\
		\end{cases}\\ 
		&\begin{cases}
			h^+=\underset{h' \in \Xi}{\arg\min} \ \mathbb{U}(\mathbb{X}_e,h'), & (\mathbb{X}_e,h) \in D' \\
		\end{cases}\\
		& C':=\{(\mathbb{U}(\mathbb{X}_e,h)-\min_{h' \in \Xi}  \mathbb{U}(\mathbb{X}_e,h')\le \delta), \\
		& D':=\{(\mathbb{U}(\mathbb{X}_e,h)-\min_{h' \in \Xi} \mathbb{U}(\mathbb{X}_e,h')\ge \delta), \\
	\end{split}
\end{equation}
	
\textbf{Theorem 2}: Consider the robot dynamic equation (\ref{eq13}) in a closed loop with the proposed hybrid feedback law (\ref{eq23}) and observer (\ref{eq20}). Then, the compact set $\mathcal{A}:=\{\mathbb{X}_d \in SE(3),\tilde{\mathcal{X}} \in SLAM_n(3) : \mathbb{X}_d = \bar{\Psi}(R_d,p_d), \tilde{\mathcal{X}}=I\}$ is globally asymptotically stable for the resulting closed-loop systems.
	
\textbf{Proof}: In accordance with Lemma 1, the proof of Theorem 2 fall into two phases.
	
\textbf{Step 1}: The second condition of (\ref{eq7}) is proven in this step. It can easily be shown $\dot{\mathbb{X}}^{-1}=-\mathbb{X}^{-1} \dot{\mathbb{X}} \mathbb{X}^{-1}$ by using the fact that $\mathbb{X}^{-1}\mathbb{X}=I_{4 \times 4}$. The end-effector motion is represented by the following kinematic model:
	
\begin{equation}\label{eq25}
	\dot{\mathbb{X}} = \mathbb{X}\left[ \begin{matrix}
		\Gamma(\omega) & v \\
		0 & 0 \\
	\end{matrix} \right] =  \mathbb{X}\mathbb{W}
\end{equation}
	
To define the end-effector pose tracking error $\mathbb{X}_e=\mathbb{X}_d^{-1}\mathbb{X}$, one can formulate the tracking error dynamics as follows.
	
\begin{equation}\label{eq26}
	\begin{split}
		& \dot{\mathbb{X}}_e = \dot{\mathbb{X}}_d^{-1}\mathbb{X}+\mathbb{X}_d^{-1}\dot{\mathbb{X}} \Rightarrow \\
		& \dot{\mathbb{X}}_e = \mathbb{X}_e(\mathbb{W}-Ad_{\mathbb{X}_e^{-1}}\mathbb{W}_d)=\mathbb{X}_e\mathbb{Y} \\
	\end{split}
\end{equation}
	
Here, the definition of $\mathbb{W}_d$ is the same as that of $\mathbb{W}$ with the desired constant angular velocity $\omega_d$ and constant linear velocity $v_d$. This paper employs the Lyapunov candidate function as follows:
	
\begin{equation}\label{eq27}
	V(\mathbb{X}_e,\tilde{\mathcal{X}},\mathbb{Y}) = \mathbb{U}(\mathbb{X}_e,h)+\mathcal{U}(\mathcal{\tilde{X}})+\frac{1}{2}\bar{\varphi}^T(\mathbb{Y})\bar{\varphi}(\mathbb{Y})
\end{equation}
	
The time derivative of the proposed Lyapunov function is given by 
	
\begin{equation}\label{eq28}
	\dot{V} = \left\langle \nabla_{\mathbb{X}_e}\mathbb{U},\mathbb{X}_e\mathbb{Y} \right\rangle_{\mathbb{X}_e}+\left\langle \nabla_{\mathcal{\tilde{X}}}\mathcal{U},\dot{\tilde{\mathcal{X}}} \right\rangle_{\tilde{\mathcal{X}}}+\bar{\varphi}^T(\mathbb{Y})\bar{\varphi}(\dot{\mathbb{Y}}).
\end{equation}
	
Therefore, by substituting Equations 24 and 27 \cite{wang2018geometric} into Equation (\ref{eq28}) and using the fact that $\dot{\mathbb{W}}_d=0$, one gets
	
\begin{equation}\label{eq29}
	\begin{split}
		& \dot{V} = -k_o\| \Upsilon((I-\tilde{\mathcal{X}}^{-1})A) \|_F^2 \\
		& + \left\langle \mathbb{X}_e^{-1}\nabla_{\mathbb{X}_e}\mathbb{U},\mathbb{Y} \right\rangle +\bar{\varphi}^T(\mathbb{Y})\bar{\varphi}(\dot{\mathbb{W}}) \\
	\end{split}
\end{equation}
	
In Equation (\ref{eq29}), $\bar{\varphi}(\dot{\mathbb{W}})$ is achieved by taking the time derivative of Equation (\ref{eq15}) since $\bar{\varphi}(\mathbb{W}) = \mathcal{Z}$. Then,
	
\begin{equation}\label{eq30}
	\dot{\mathcal{Z}} = J\ddot{q}+\dot{J}\dot{q},
\end{equation}
and replacing $\ddot{q}$ by Equation (\ref{eq13}), one gets
	
\begin{equation}\label{eq31}
	\dot{\mathcal{Z}} = JM^{-1}(q)(\tau-N(q,\dot{q}))+\dot{J}\dot{q}.
\end{equation}
	
The following equation is the result of applying the proposed hybrid feedback law ($\tau^*$) to Equation (\ref{eq31}):
	
\begin{equation}\label{eq32}
	\dot{\mathcal{Z}} = -\bar{\varphi}(\mathbb{X}_e^{-1}\nabla_{\mathbb{X}_e}\mathbb{U}+G_d\mathbb{Y})
\end{equation}
	
Consequently, Equation (\ref{eq29}) is simplified to
	
\begin{equation}\label{eq33}
	\begin{split}
		& \dot{V} = -k_o\| \Upsilon((I-\tilde{\mathcal{X}}^{-1})A) \|_F^2-g_d\bar{\varphi}^T(\mathbb{Y})\bar{\varphi}(\mathbb{Y}) \\
		& + \left\langle \mathbb{X}_e^{-1}\nabla_{\mathbb{X}_e}\mathbb{U},\mathbb{Y} \right\rangle - \bar{\varphi}^T(\mathbb{Y})\bar{\varphi}(\mathbb{X}_e^{-1}\nabla_{\mathbb{X}_e}\mathbb{U}), \Rightarrow\\
		&  \dot{V} = -k_o\| \Upsilon((I-\tilde{\mathcal{X}}^{-1})A) \|_F^2-g_d\bar{\varphi}^T(\mathbb{Y})\bar{\varphi}(\mathbb{Y}) \le 0. \\
	\end{split}
\end{equation}
	
As a result, it follows from (\ref{eq33}) that both the tracking error and estimation error are globally bounded, hence; $\ddot{V}$ is also globally bounded. By invoking Barbalat's lemma, it can be deduced that $\lim_{t \to +\infty} \dot{V}=0$; therefore, $\lim_{t \to +\infty} \mathbb{X} \rightarrow \mathbb{X}_d$ and $\lim_{t \to +\infty} \mathcal{X} \rightarrow \hat{\mathcal{X}}$. \\
	
\textbf{Step 2}: This step provides proof for the third condition of (\ref{eq7}). Due to the existence of switching variables $(h,q)$, it is necessary to test variation in $V(\mathbb{X}_e,\tilde{\mathcal{X}},\mathbb{Y})$ to guarantee that the Lyapunov function remains negative during jumps. Given the last condition of (\ref{eq7}), the variation in $V$ among jumps is defined by
	
\begin{equation}\label{eq34}
	\begin{split}
		& V^+(\mathbb{X}_e,\tilde{\mathcal{X}},\mathbb{Y})-V(\mathbb{X}_e,\tilde{\mathcal{X}},\mathbb{Y})= \\
		& \mathbb{U}(\mathbb{X}_e,h^+)+\mathcal{U}(\mathcal{\tilde{X^+}})-\mathbb{U}(\mathbb{X}_e,h)-\mathcal{U}(\mathcal{\tilde{X}})= \\
		& \mathcal{U}(\mathcal{\tilde{X}}_q)-\mathcal{U}(\mathcal{\tilde{X}})+\mathbb{U}(\mathbb{X}_e,h^+)-\mathbb{U}(\mathbb{X}_e,h)\\
	\end{split}
\end{equation}
	
From (\ref{eq24}) and (\ref{eq20}), one can determine that
	
\begin{equation}\label{eq35}
	\begin{split}
		& \min_{\tilde{\mathcal{X}_q}\in \mathcal{Q}} \mathcal{U}(\tilde{\mathcal{X}_q})-\mathcal{U}(\tilde{\mathcal{X}}) \le -\delta, \\
		& \min_{h' \in \Xi} \mathbb{U}(\mathbb{X}_e,h') - \mathbb{U}(\mathbb{X}_e,h) \le -\delta. \\
	\end{split}
\end{equation}
	
\begin{figure}
	\centering
	\includegraphics[width=1\linewidth]{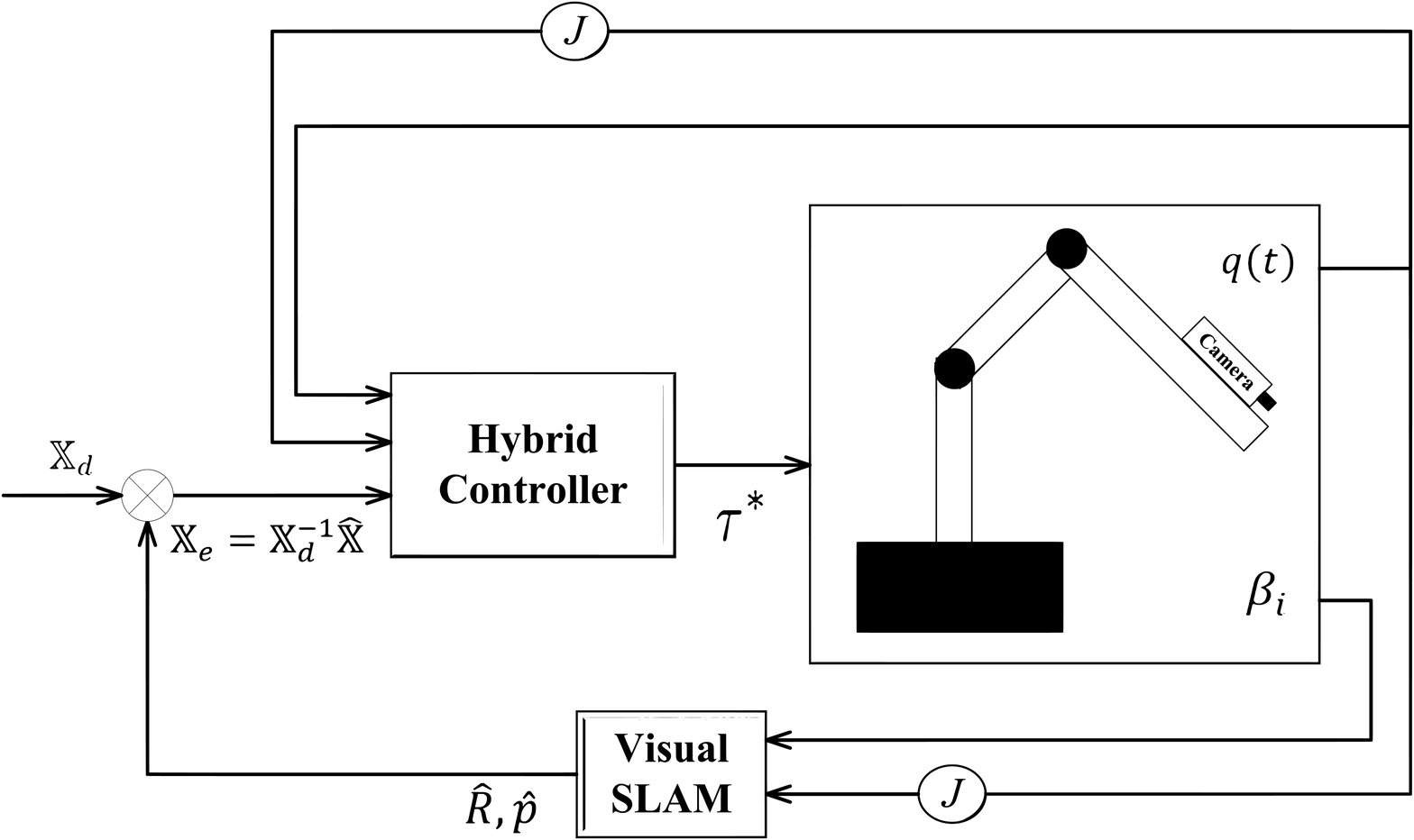}
	\caption{Block diagram of the proposed scheme.}
	\label{fig1}
\end{figure}

Accordingly, from Lemma 1, one can easily derive that set $\mathcal{A}$ is globally asymptotically stable. $\square$\\
It is worth noting that the estimated pose $\bar{\Psi}(\hat{R},\hat{p})$ can be substituted for the true pose $\bar{\Psi}(R,p)$ in the proposed feedback law without invalidating the stability proof since $\lim_{t \to +\infty} \mathcal{X} \rightarrow \hat{\mathcal{X}}$. The block diagram of the proposed VSLAM-based control structure is depicted in Figure (\ref{fig1}), and the salient features of the proposed method are 1) its simple structure, 2) global stability, and 3) light computational burden.
	
\section{Simulation Results}
\begin{figure}
	\centering
	\includegraphics[width=1\linewidth]{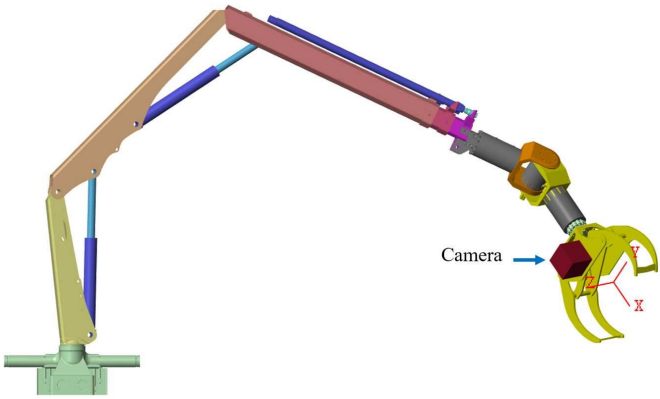}
	\caption{6-DOF manipulator arm utilized to validate the proposed structure.}
	\label{fig2}
\end{figure}

In this section, a numerical simulation study is performed to assess the performance of the proposed VSLAM-based control structure given by (\ref{eq23}). This study uses a 6-DOF manipulator with a long arm, illustrated in Figure (\ref{fig2}). A detailed description of and supplementary material on this manipulator can be found in \cite{petrovic2022mathematical}. The observer gain and controller gain were determined as follows through trial and error until a satisfactory performance was obtained:
	
\begin{equation*}
K = 100 I_{n \times n}, \quad G = 200 I_{4 \times 4}, \quad G_d = 5 I_{6 \times 6}
\end{equation*}
	
\begin{figure}
	\centering
	\includegraphics[width=1\linewidth]{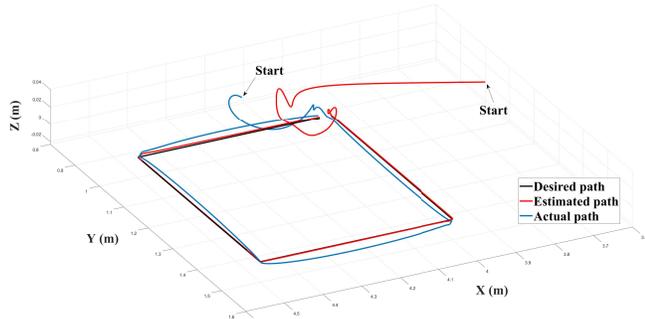}
	\caption{End-effector trajectory, desired trajectory, and estimated trajectory in 3D space.}
	\label{fig3}
\end{figure}

\begin{figure}
	\centering
	\includegraphics[width=1\linewidth]{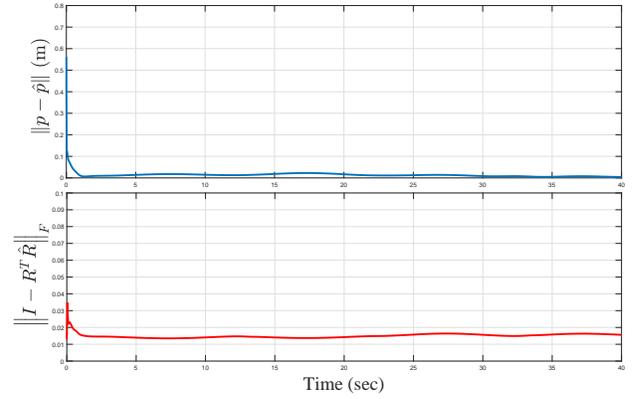}
	\caption{Position and attitude estimation errors.}
	\label{fig4}
\end{figure}

\begin{figure}
	\centering
	\includegraphics[width=1\linewidth]{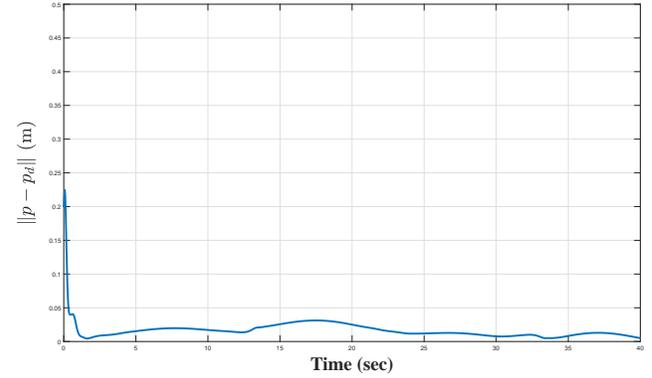}
	\caption{Position tracking error of the end effector.}
	\label{fig5}
\end{figure}

\begin{figure}
	\centering
	\includegraphics[width=1\linewidth]{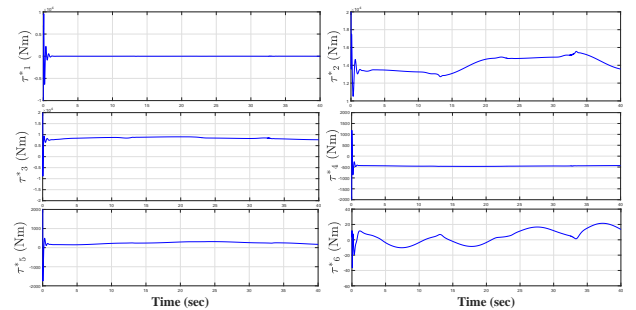}
	\caption{Control input torque.}
	\label{fig6}
\end{figure}

Here, the task of the robot is to draw a square with sides of 50 cm to demonstrate the robustness of the proposed observer against initial conditions. The initial position of the observer is chosen as $[3.6 \ 1 \ 0]$, which is different from the true value. The simulation results in Figures (\ref{fig3}-\ref{fig6}) are achieved by applying the proposed VSLAM-based control scheme to the 6-DOF manipulator with the long arm. The desired trajectory, actual trajectory, and trajectory estimated by VSLAM are depicted in Figure (\ref{fig3}). This Figure shows that the proposed controller has high accuracy in path tracking. Figure (\ref{fig4}) demonstrates the Euclidean norm of position estimation error and Frobenius norm of attitude estimation error. The end-effector position tracking error versus time is shown in Figure (\ref{fig5}). It can be deduced from Figures (\ref{fig4}-\ref{fig5}) that a tracking error of approximately $2\%$ is obtained despite the existence of $1\%$ error in the estimated position. The torque produced by the introduced VSLAM-based control structure is illustrated in Figure (\ref{fig6}). This figure confirms that the torque produced by the proposed controller is realizable and applicable.
	
\section{Conclusion}
This paper investigated the problem of designing a globally stable vision-based controller for robot manipulators. To address this problem, a novel control structure was introduced in which a visual SLAM observer was employed for estimating the robot pose. Based on feedback information provided by the VSLAM observer, a new hybrid feedback law was directly designed in $SE(3)$. The proposed hybrid controller was derived by taking the gradient of a potential function defined in $SE(3)$. The global asymptotic stability of the proposed VSLAM-based control structure was proven with the help of the Lyapunov stability theorem. Finally, the proposed control scheme was tested on a 6-DOF robot manipulator to demonstrate its accuracy and efficiency.
	
\appendices
\section{Useful Properties of $SLAM_n(3)$ and $SE(3)$}
This subsection presents some useful identities, properties, and maps related to the matrix Lie groups $SLAM_n(3)$ and $SE(3)$:
	
\begin{equation*}
	\begin{split}
		& \Upsilon(\mathcal{X}B)=\Upsilon(\mathcal{X}^{-T}B), \quad (a) \\
		& \left\langle \mathcal{V},B \right\rangle=\left\langle \mathcal{V},\Upsilon(B) \right\rangle=\left\langle \Upsilon(B),\mathcal{V} \right\rangle, \quad (b)\\
		&\text{tr}(ABCD)=\text{tr}(CDAB)=\text{tr}(DABC), \quad (c)\\
		&\text{tr}(\mathcal{X}^T\mathcal{X}\Upsilon(B)\Upsilon(B)^T)=\text{tr}(\Upsilon(B)\Upsilon(B)^T), \quad (d) \\
		& \frac{\partial \text{tr}(A\mathcal{X}B\mathcal{X}^TC)}{\partial \text{tr}(\mathcal{X})}=B\mathcal{X}^TCA+B^T\mathcal{X}^TA^TC^T \ (e) \\
	\end{split}
\end{equation*}
	
\bibliographystyle{ieeetran}
\bibliography{mybibfile}
	
\begin{IEEEbiography}[{\includegraphics[width=1in,height=1.25in,clip,keepaspectratio]{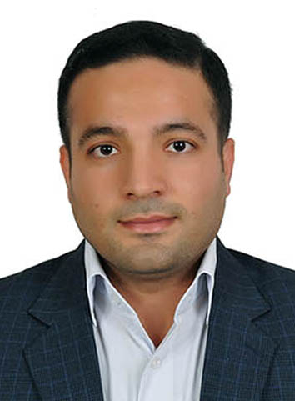}}]{Seyed Hamed Hashemi} received a B.Sc. in electrical engineering from Babol Noshirvani University of Technology (BNUT), Iran, in 2014; an M.Sc. in electrical engineering-control from Shahrood University of Technology (SUT), Iran, in 2017; and a Ph.D. in electrical engineering-control from Ferdowsi University of Mashhad (FUM), Iran, in 2021. He is currently a postdoctoral research fellow with the Automation Technology and Mechanical Engineering Unit in the Faculty of Engineering and Natural Sciences, Tampere University. His research interests include topological constraints in control systems, the control of hybrid systems, simultaneous localization and mapping, and estimation theory.
\end{IEEEbiography}
	
\begin{IEEEbiography}[{\includegraphics[width=1in,height=1.25in,clip,keepaspectratio]{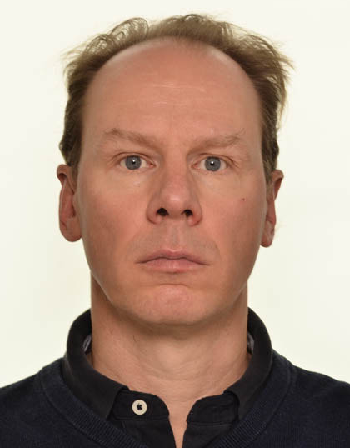}}]{Jouni Mattila} received his M.Sc. and Ph.D. in automation engineering, both from the Tampere University of Technology, Tampere, Finland, in 1995 and 2000, respectively. He is currently a professor of machine automation with the Automation Technology and Mechanical Engineering Unit, Tampere University. His research interests include machine automation, nonlinear model-based control of robotic manipulators, and energy-efficient control of heavy-duty mobile manipulators.
\end{IEEEbiography}
	
\end{document}